\documentclass[a4paper,11pt]{article}
\usepackage{pos}

\usepackage[utf8]{inputenc}
\usepackage{braket}

\newcommand{\tf}{t_{s}}
\newcommand{\tc}{t_{\rm ins}}

\newcommand{\OJ}{\mathcal{J}}
\newcommand{\OO}{\mathcal{O}}
\newcommand{\JN}{\OJ_N}

\newcommand{\JNd}{\OJ^\dagger_N}

\newcommand{\JNpi}{\OJ_{N\pi}}

\newcommand{\JNpid}{\OJ^\dagger_{N\pi}}
\newcommand{\NN}{\braket{\JN\JNd}}
\newcommand{\NNpi}{\braket{\JN\JNpid}}
\newcommand{\NpiN}{\braket{\JNpi\JNd}}
\newcommand{\NpiNpi}{\braket{\JNpi\JNpid}}
\newcommand{\NJN}{\braket{\JN\OO\JNd}}
\newcommand{\NJNpi}{\braket{\JN\OO\JNpid}}
\newcommand{\NpiJN}{\braket{\JNpi\,\OO\,\JNd}}
\newcommand{\NpiJNpi}{\braket{\JNpi\,\OO\,\JNpid}}

\newcommand{\Cdpt}{C}

\newcommand{\rcite}[1]{\cite{#1}}

\newcommand{\refrefs}[1]{Refs.~\rcite{#1}}
\newcommand{\eref}[1]{Eq.~(\ref{#1})}

\newcommand{\tref}[1]{Table~\ref{#1}}
\newcommand{\fref}[1]{Fig.~\ref{#1}}

\title{Investigation of $\pi N$ contributions to nucleon matrix elements}

\author[a,b]{Constantia Alexandrou}
\author[b]{Giannis Koutsou}
\author*[a]{Yan Li}
\author[c]{Marcus Petschlies}
\author[b]{Ferenc Pittler}

\affiliation[a]{Department of Physics,
University of Cyprus, P.O. Box 20537, 1678 Nicosia, Cyprus}

\affiliation[b]{Computation-based Science and Technology Research Center,
The Cyprus Institute, 20 Kavafi Str., Nicosia 2121, Cyprus}

\affiliation[c]{HISKP (Theory), Rheinische
Friedrich-Wilhelms-Universit{\"a}t Bonn, Nu{\ss}allee 14-16, 53115
Bonn, Germany}

\emailAdd{li.yan@ucy.ac.cy}

\abstract{We investigate an improved method to extract nucleon matrix elements from lattice 3-point functions using a generalized eigenvalue problem (GEVP) with nucleon and pion-nucleon interpolating fields. Our method avoids the computation of the costly three-point functions that have pion-nucleon interpolators at both source and sink. We demonstrate that excited state contamination from $N\pi$ is minimized in nucleon matrix elements of the scalar, vector, pseudoscalar, axial, and tensor currents and discuss our results based on a physical-point ensemble with a pion mass value of 131~MeV.
We find that the GEVP is most significant for the isovector pseudoscalar and axial currents.
}

\FullConference{The 41st International Symposium on Lattice Field Theory (LATTICE2024)\\
 28 July - 3 August 2024\\
Liverpool, UK\\}

\begin{document}
\maketitle

\section{Introduction}
 Extracting nucleon matrix elements from three-point correlators requires a careful analysis in order to minimize contributions from excited states. Since excited states may contribute with different weights in two- and three-point correlators, the study of excited states needs to be carried out for each type of current. 
Nucleon matrix elements, $\braket{N|\OO|N}$ are extracted from the ratio of three-point to two-point functions in the asymptotic time limit:
\begin{align}\label{eq:ratio}
    \frac{\braket{\OJ_N(t_{\text{s}})\OO(t_{\text{ins}})\bar{\OJ}_{N}(0)}}{\braket{\OJ_N(t_{\text{s}})\bar{\OJ}_{N}(0)}} \xrightarrow[t_{\rm s}- t_{\rm sink}\to\infty]{\text{$t_{\rm ins} \rightarrow \infty$ }} \braket{N|\OO|N} \,.
\end{align}
Since statistical errors increase exponentially with the time separation, in practice we are limited to separations that often still exhibit a time dependence that comes from contributions of excited states.
A dominant contribution to three-point functions may come from the lowest energy nucleon-pion  state  $N\pi$ with the quantum numbers of the nucleon. For instance, in Refs.~\cite{Gupta:2021ahb,Gupta:2023cvo}, it was demonstrated that the discrepancy in the nucleon $\sigma$-term, $\sigma_{\pi N}$, between lattice and phenomenology can be resolved if one accounts for an excited state energy close to the  $N\pi$ or $N\pi\pi$ state. %
Moreover, in Ref.~\cite{Barca:2022uhi}  significant improvements using $N\pi$ operators are found in the isovector pseudoscalar and axial channels. That analysis was performed using a gauge  ensemble with pion mass $m_\pi=429$~MeV.
In this work, we use as basis the nucleon and $N\pi$ interpolators to study nucleon three-point correlators of the scalar, vector, pseudoscalar, axial-vector and tensor bilinear operators and a gauge ensemble simulated at a physical value of the pion mass.

\section{Lattice setup}
\label{sec:setup}

We perform the analysis using a gauge  ensemble    simulated using twisted mass clover-improved fermions and light quarks $(N_f=2)$ with mass tuned to approximately their physical values. The parameters are given in \tref{tab:ens}.
In  Ref.~\cite{Alexandrou:2024tin} we provide details using in addition a heavier pion mass for the analysis that follows.

\begin{table}[h!]
    \caption{Parameters of the gauge ensemble used in this work. Further details are given in 
    \refrefs{ETM:2015ned,Alexandrou:2018egz,ExtendedTwistedMass:2021gbo}. The right-most column gives the number of gauge configurations $N_{\text{cfg}}$ employed in this analysis.}\label{tab:ens}
    \centering
    \renewcommand\arraystretch{1.5}
    \begin{tabular}{ccccccccc}
      \hline\hline
      Ensembles & Flavors (N\textsubscript{f}) & $N_L^3\times N_T$ & $a$ [fm] & $L$ [fm] & $m_\pi$ [MeV] & $m_N$ [MeV] & $N_{\rm cfg}$ \\ \hline
      cA2.09.48 & 2 & $48^3 \times 96$ & 0.0938 & 4.50 & 131  & 931(3) & 1228 \\\hline\hline
    \end{tabular}
\end{table}

We consider single nucleon $\OJ_N$ and nucleon-pion $\OJ_{N\pi}$ interpolating fields at various momenta. 
We compute the two-point functions $\NN$, $\NNpi$, $\NpiN$, $\NpiNpi$, and the three-point functions $\NJN$, $\NJNpi$, $\NpiJN$.
We omit the $\NpiJNpi$ three-point function,  which requires substantially more computational resources.
We compute both connected and disconnected topologies for all computed 2-point and 3-point functions.
We note that due to the breaking of the isospin symmetry at finite lattice spacing $a$ exhibited by the twisted-mass fermion action, the three-point functions with the disconnected quark loop contributions are nonzero also for isovector insertion operators. They are therefore also computed here. 

\section{Reference time dependence of eigenstates of the generalized eigenvalue problem (GEVP)}
\begin{figure*}[!ht]
  \centering
  \includegraphics[width=\textwidth]{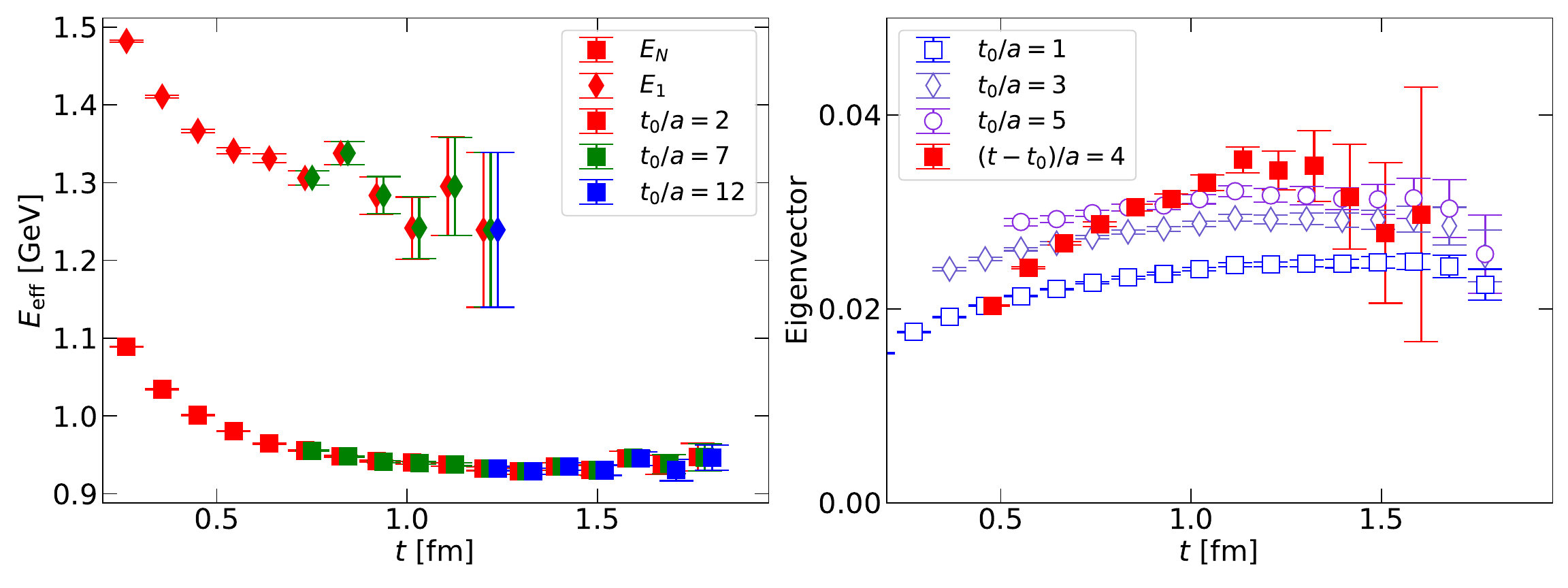}
  \caption{Example of the $t_0$-dependence of the effective energies (left) and eigenvectors (right) for  zero total momentum. 
The eigenvector component used is $|v_{0,N\pi}/v_{0,N}|$.}\label{fig:t0dep}
\end{figure*}

The GEVP equation is given by 
\begin{align}\label{eq:GEVP}
  \sum_{k}\Cdpt_{jk}(t)\,v_{nk}(t,t_0) = \lambda_n(t,t_0)\sum_{k}\Cdpt_{jk}(t_0)\,v_{nk}(t,t_0) \,,
\end{align}
where $C_{jk}(t)$ is the correlation matrix of the interpolator basis, $\nu_{jk}(t,t_0)$ are eigenvectors and $\lambda_n(t,t+0)$ the eigenvalues. Both eigenvectors and eigenvalues depend on $t$ and a reference time slice $t_0$, which is typically fixed to a small value in most spectral analyses of nucleon correlators.
However, as discussed in Ref.~\cite{Blossier:2009kd}, the use of a small $t_0$ can bring large systematic errors. 
In \fref{fig:t0dep}, we show results for the $t_0$ dependence of both effective energies and eigenvectors.
The effective energies built of eigenvalues show little dependence on $t_0$ reflected by the overlapping of points of different colors, justifying the use of a small fixed $t_0$.
However, the eigenvectors show stronger $t_0$-dependence reflected by the open symbols. Therefore, in this work, we choose to fix $t-t_0$ and look for the stability when $t$ and $t_0$ increase together.

\section{Optimal combination of three-point functions using GEVP}
For the $N$-$N\pi$ system,  we  obtain  an improved nucleon interpolator 
\begin{align}
  \tilde{\OJ}_{N}:=v_{0,N}{\OJ}_{N} + v_{0,N\pi}{\OJ}_{N\pi}
\end{align}
 by solving the GEVP, which has a larger overlap  with the nucleon ground state compared to $\OJ_N$. Replacing $\OJ_N$ in \eref{eq:ratio} with $\tilde{\OJ_N}$, we obtain for the numerator:  
\begin{align}
  I&=v_{0,N}v_{0,N}^*\braket{\OJ_N(t_{\text{s}})\OO(t_{\text{ins}})\bar{\OJ}_{N}(0)} + v_{0,N}v_{0,N\pi}^* \braket{\OJ_N(t_{\text{s}})\OO(t_{\text{ins}})\bar{\OJ}_{N\pi}(0)} \nonumber\\
  +& v_{0,N\pi}v_{0,N}^* \braket{\OJ_{N\pi}(t_{\text{s}})\OO(t_{\text{ins}})\bar{\OJ}_{N}(0)} + v_{0,N\pi}v_{0,N\pi}^* \braket{\OJ_{N\pi}(t_{\text{s}})\OO(t_{\text{ins}})\bar{\OJ}_{N\pi}(0)} \,.
\end{align}
The last term in the above expression is  computationally more demanding and it is not considered in this work. Instead, we consider an application of the GEVP by introducing additional weight factors:
\begin{align}
  I_d&=d_{N,N}\,v_{0,N}v_{0,N}^*\braket{\OJ_N(t_{\text{s}})\OO(t_{\text{ins}})\bar{\OJ}_{N}(0)} + d_{N,N\pi}\,v_{0,N}v_{0,N\pi}^* \braket{\OJ_N(t_{\text{s}})\OO(t_{\text{ins}})\bar{\OJ}_{N\pi}(0)} \nonumber\\
  +& d_{N\pi,N}\,v_{0,N\pi}v_{0,N}^* \braket{\OJ_{N\pi}(t_{\text{s}})\OO(t_{\text{ins}})\bar{\OJ}_{N}(0)} + d_{N\pi,N\pi}\,v_{0,N\pi}v_{0,N\pi}^* \braket{\OJ_{N\pi}(t_{\text{s}})\OO(t_{\text{ins}})\bar{\OJ}_{N\pi}(0)} \,.
\end{align}
%
%
%
The quantity  $I$ is by construction the combination that suppresses the contamination from $\braket{N|\OO|N\pi}$, $\braket{N\pi|\OO|N}$ and $\braket{N\pi|\OO|N\pi}$ and thus isolates the desired matrix element $\braket{N|\OO|N}$. The new quantity $I_d$, that does not include the  diagonal matrix element with the ${\cal J}_{N\pi}$ interpolator by forcing $d_{N\pi,N\pi}=0$,  in general is not an optimal combination. However, we make the key observation that when the source-sink and source-insertion time separations increase, the contamination for the diagonal matrix element $\braket{N\pi|\OO|N\pi}$ decreases faster than  the off-diagonal terms.  Therefore, requiring the elimination of the off-diagonal contaminations while keeping only the terms with $d_{N,N}$, $d_{N,N\pi}$, and $d_{N\pi,N}$  we find
\begin{align}
  d_{N,N} = 1 - W^*\,W \,,\quad d_{N,N\pi}=1+W^*\,,\quad d_{N\pi,N}=1+W \,,
\end{align}
with
\begin{align}
  W = \frac{1}{v_{0,N}\,[v^{-1}]_{N,0}} -1 \,
\end{align}
where $[v^{-1}]$ is the inverse  of the eigenvector matrix.
Therefore, the weights $d$ used to construct $I_d$ can be determined from the GEVP, and are independent of the insertion operator of the 3-point function.

\section{Lattice results}
We apply this approach   to the nucleon matrix elements of the isoscalar and isovector scalar, pseudoscalar, vector, axial, and tensor bilinear operators.
We obtain results using a range of kinematical setups, including having momentum in the sink and in addition to having momentum at the source.
Comparing the results obtained with  the GEVP optimized operators to those extracted with  only using  $\OJ_N$, we do not observe any significant improvement for the majority  of cases. For more details see Ref.~\cite{Alexandrou:2024tin}.
%
%
%
In the following, we present results for the case with a scalar current where we do not observe an improvement, as well as for the cases where we do observe an improvement, such as for the isovector pseudoscalar and axial currents.

\begin{figure}[!ht]
  \centering
  \includegraphics[width=11cm]{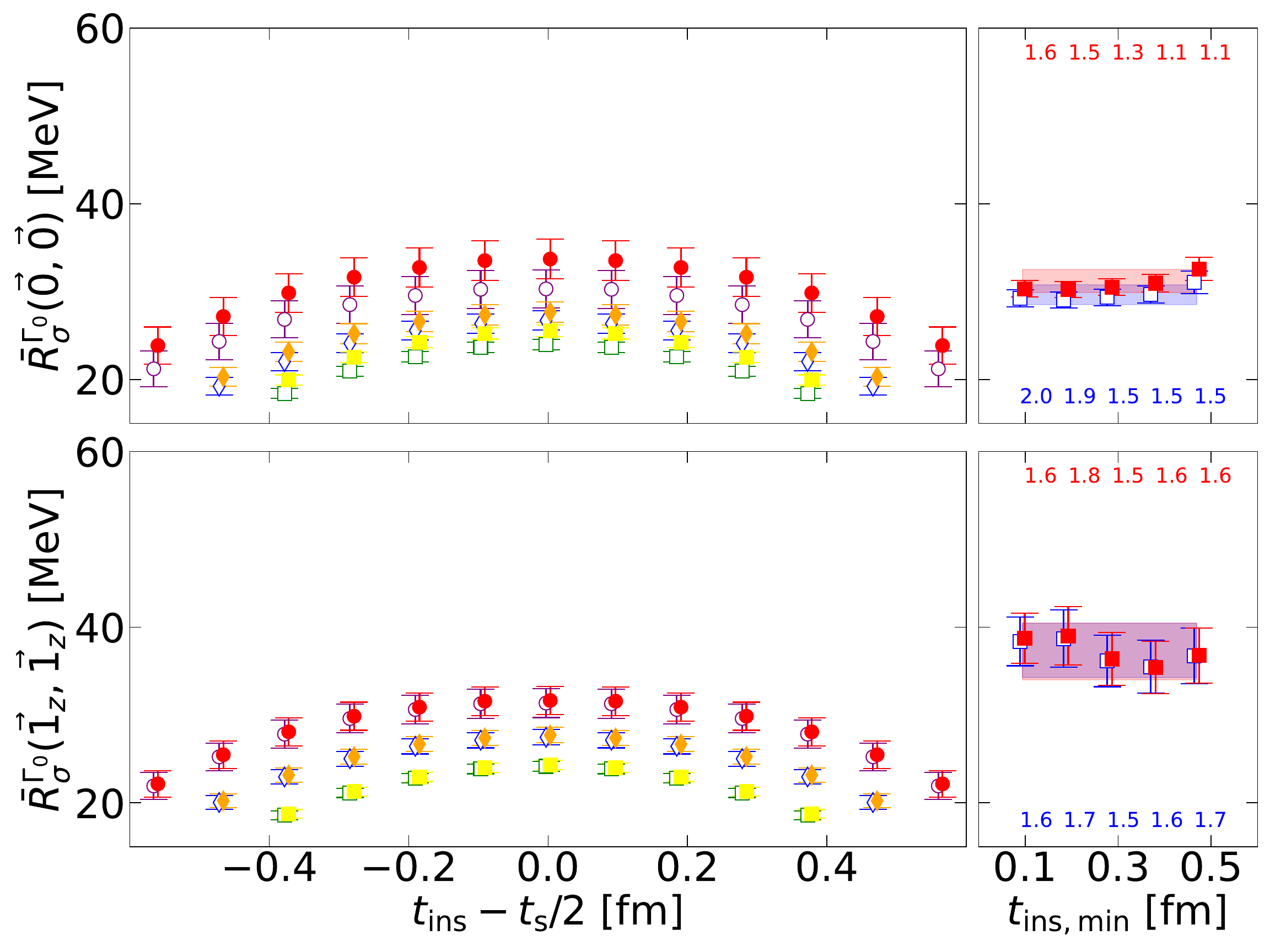}
  \caption{%
  Ratios that yield $\sigma_{\pi N}$ versus $t_{\rm ins}-t_s/2$ (left) and the results of two-state fits to two- and three-point functions versus the smallest insertion time $t_{\rm ins, min}$  in the fit (right). We compare between results with (filled symbols) and without (open symbols) the use of operators improved via the GEVP.
  The blue (without GEVP) and red (with GEVP) bands are the results from the model average of the two-state fit to the corresponding results.
  The reduced $\chi^2$ values for each fit are given in the right panel with the corresponding color to the bands. We show results for both nucleon at rest (top) or with one unit of momentum (bottom). 
  }\label{fig:3pt_raw_sep_S_48}
\end{figure}

In \fref{fig:3pt_raw_sep_S_48}, we show results for two ratios that yield $\sigma_{\pi N}$. 
The ratio for the first (second) row is constructed using in the nucleon rest-frame (moving-frame). We do not observe a significant improvement for either case when including GEVP-optimized operators. We therefore conclude that the contamination observed in the $\sigma_{\pi N}$ case is unlikely from the lowest $N\pi$ states under consideration.

\begin{figure}[!ht]
  \centering
  \includegraphics[width=11.cm]{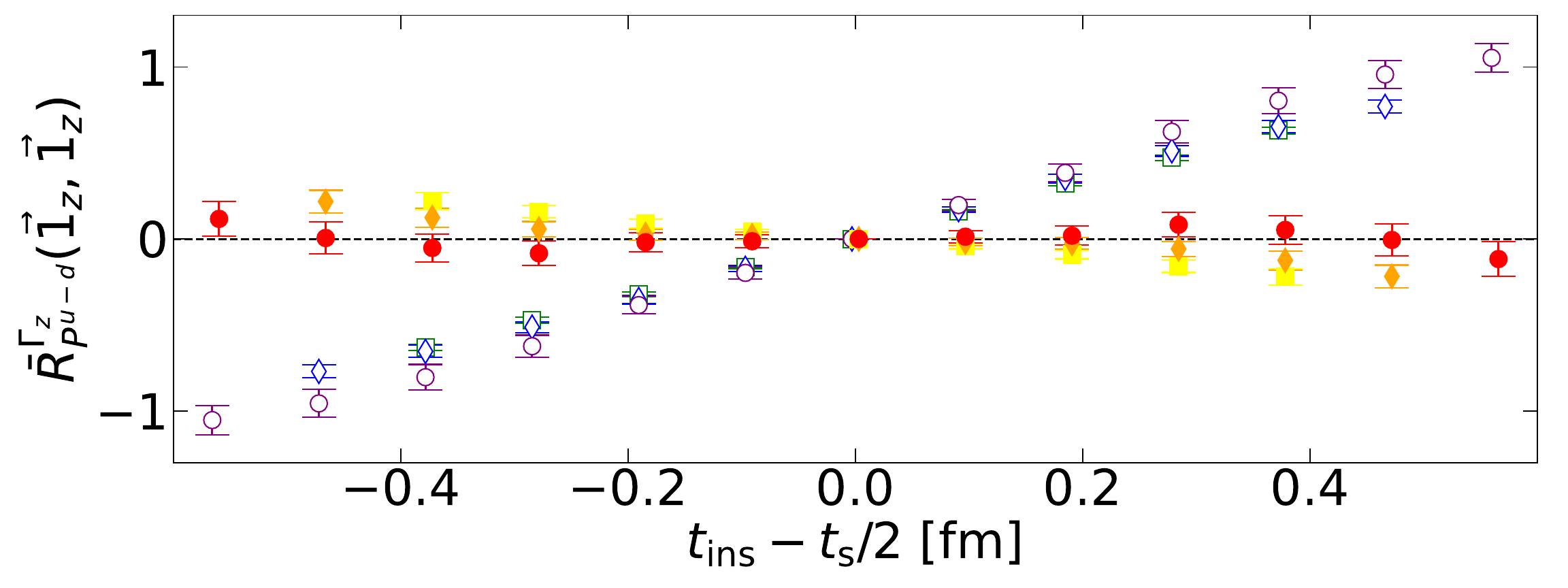}
  \caption{
    Ratios of the pseudoscalar insertion operator in a setup that should yield zero due to parity symmetry. The rest of the notation is the same as in the left panel of \fref{fig:3pt_raw_sep_S_48}.
  }\label{fig:3pt_raw_sep_J_48}
\end{figure}

In \fref{fig:3pt_raw_sep_J_48}, we show results for the ratio with the pseudoscalar insertion operator that should yield  zero due to parity symmetry.
The ratio is exactly zero at the mid-point $\tc=\tf/2$ since the data are  symmetrized.
As can be seen, when using the single nucleon operator, the ratio is non-zero for all time separations except at the mid-point, indicating significant excited state contamination.
When using the improved operator obtained from GEVP, the ratio is consistent with zero  (or within two standard deviations at large time separations) for all of $\tc-\tf/2$values.
Therefore, we conclude that the dominant excited state contamination observed in this case comes from the $N\pi$ states under consideration, and is successfully removed when employing our approach.

\begin{figure}[h!]
  \centering
  \includegraphics[width=11cm]{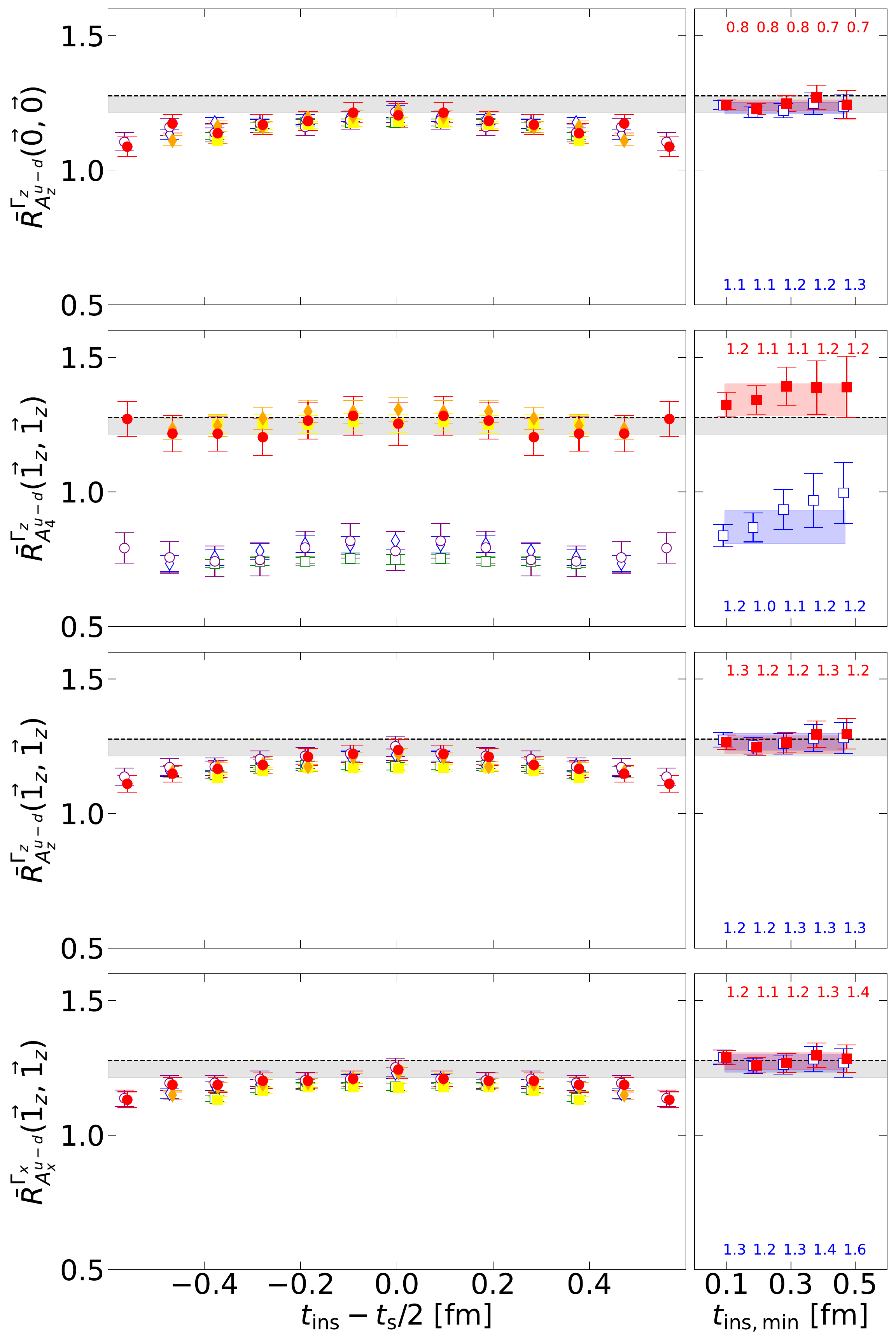}
  \caption{
    As in \fref{fig:3pt_raw_sep_S_48} for ratios that yield $g_A^{u-d}$.
    The grey band in each panel is from Ref.~\cite{Alexandrou:2023qbg}.
  }\label{fig:3pt_raw_sep_AJ_48}
\end{figure}
\begin{figure}[h!]
  \centering
  \includegraphics[width=11cm]{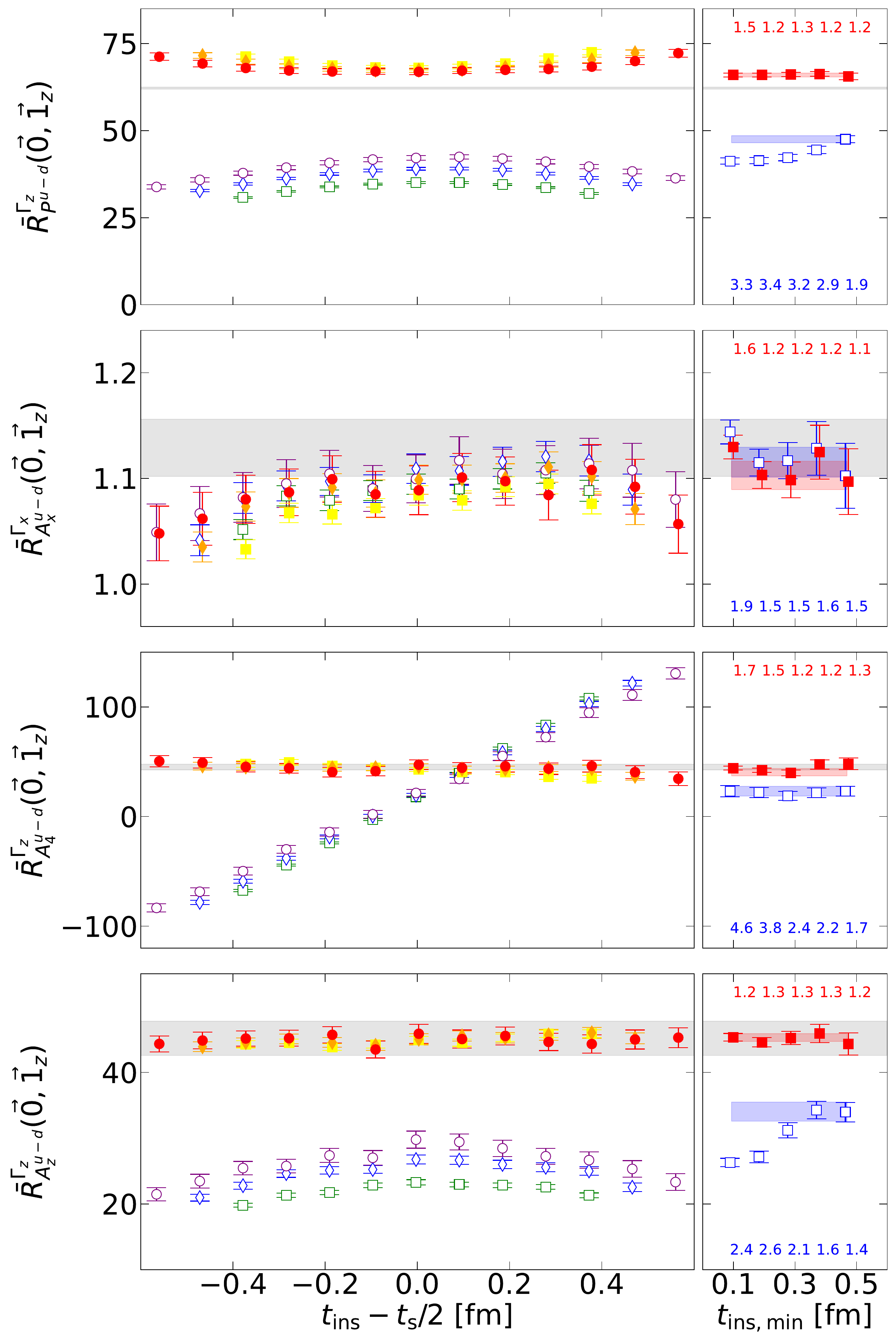}
  \caption{As in \fref{fig:3pt_raw_sep_S_48} for ratios that yield $G_5^{u-d}$ (first row), $G_A^{u-d}$ (second row) and $G_P^{u-d}$ (third and last rows) at one-unit of transfer momentum.
  The grey band in each panel is from Ref.~\cite{Alexandrou:2023qbg}.
  }\label{fig:3pt_raw_sep_5APP_48}
\end{figure}
In \fref{fig:3pt_raw_sep_AJ_48}, we show results for four ratios that yield the isovector axial charge $g_A^{u-d}$ using nucleon interpolator for both rest and moving frame and with different components of the axial insertion operator.
The grey band in each panel is taken from Ref.~\cite{Alexandrou:2023qbg}, where the continuum limit was taken using physical point ensembles with $a=0.07957(13)$ fm, $a=0.06821(13)$ fm, and $a=0.05692(12)$ fm and after a  dedicated excited state analysis using up to 10 sink-source separations depending on the ensemble.
%
%
%
%
When using the single nucleon interpolating operator, we agree on all ratios used to construct $g_A$ except the timelike axial case (second row) as illustrated in \fref{fig:3pt_raw_sep_AJ_48}. The case for the second row, which involves the temporal component of the axial charge, agrees only when the GEVP-improved interpolating operator is used. Since the data presented here are from a different ensemble with different lattice artifacts, the comparison suggests that cutoff effects are small and thus enhances the reliability and robustness of the extraction of the axial charge.

In \fref{fig:3pt_raw_sep_5APP_48}, we show results for four ratios that yield, from top to bottom, the isovector pseudoscalar $G_5^{u-d}$, axial $G_A^{u-d}$, and two components of the induced pseudoscalar $G_P^{u-d}$ form factors at one unit of momentum transfer, respectively. As can be seen,  all matrix elements, except for the one from which  $G_A^{u-d}$ is extracted, are  improved  when using the GEVP, as is evident by the suppression of the time dependence of the corresponding thre-point functions. This behavior indicates that  $N\pi$ is the dominant contribution to the excited state contamination.
After applying  GEVP, our results for $G_5^{u-d}$ approach the value found in Ref.~\cite{Alexandrou:2023qbg}. We consider the remaining disagreement to be a cutoff effect as also suggested in Ref.~\cite{Alexandrou:2023qbg}.
We should also point out that, as mentioned in Sec.~\ref{sec:setup}, our current analysis for isovector quantities includes disconnected quark  loop contributions, which should vanish at the continuum limit. Including these contributions, which were not included in Ref.~\cite{Alexandrou:2023qbg}, where large cutoff effects were observed in $G_P^{u-d}$, can alter the $a^2$-dependence yielding a milder dependence on the lattice spacing.   This  explains the agreement of our results at a single lattice spacing with the continuum limit of Ref.~\cite{Alexandrou:2023qbg}.
%

\section{Conclusions}
We use a basis of single nucleon and pion-nucleon interpolating fields  and  GEVP to compute
the three-point and two-point functions for the complete set of bilinear operators.
We analyze the dependence of the eigenvectors of the GEVP on the  reference time slice $t_0$, which is found to be significant.
We develop a new method that allows us to suppress contamination using the off-diagonal three-point functions, while avoiding the computation of the costly three-point function which includes pion-nucleon interpolating field in both source and sink.
%
%
An important new component is the computation of  disconnected quark loops for isovector   current operators. We find significant reduction of  cutoff effects for the isovector pseudoscalar and induced pseudoscalar form factors.
We observe significant improvement when using  GEVP for the  nucleon matrix elements for the pseudoscalar or axial  operators.
In other cases, including the nucleon $\sigma$-term $\sigma_{\pi N}$, we find no improvement indicating that the contamination is unlikely to come from the lowest $N\pi$ state.

\section*{Acknowledgements}
We thank all members of the ETM collaboration for a most conducive cooperation.
We would like to thank Lorenzo Barca, Marilena Panagiotou and Rainer Sommer for useful discussions and suggestions.
%
We acknowledge computing time granted on Piz Daint at Centro Svizzero di Calcolo Scientifico (CSCS) via the project with id s1174, JUWELS Booster at the J\"{u}lich Supercomputing Centre (JSC) via the project with id pines, and Cyclone at the Cyprus institute (CYI) via the project with ids P061, P146 and pro22a10951.
Y.L. is supported by the Excellence Hub project "Unraveling the 3D parton structure of the nucleon with lattice QCD (3D-nucleon)" id EXCELLENCE/0421/0043 co-financed by the European Regional Development Fund and the Republic of Cyprus through the Research and Innovation Foundation.
F.P. acknowledges financial support by the Cyprus Research and
Innovation foundation Excellence Hub project NiceQuarks under contract with number
 EXCELLENCE/0421/0195. 
C.A. and G. K. acknowledge partial support from the European Joint Doctorate AQTIVATE that received funding from the European Union’s research and innovation programme under the Marie Sklodowska-Curie Doctoral Networks action under the Grant Agreement No 101072344.
M.P. acknowledges support by the Sino-German collaborative research center CRC 110.

\bibliographystyle{JHEP}
\bibliography{refs}

\end{document}